# Thermoelectric Properties of Nanoscale three dimensional Si Phononic Crystal

Lina Yang[1], Nuo Yang[2*], and Baowen Li[1,3,4,5*]

[1]Department of Physics and Centre for Computational Science and Engineering, National University of Singapore, Singapore 117542, Singapore

[2]School of Energy and Power Engineering, Huazhong University of Science and Technology (HUST), Wuhan 430074, People's Republic of China

[3]Graphene Research Center, National University of Singapore, Singapore 117546, Singapore

[4]NUS Graduate School for Integrative Sciences and Engineering, National University of Singapore, Singapore 117456, Singapore

[5]Center for Phononics and Thermal Energy Science, School of Physics Science and Engineering, Tongji University, 200092 Shanghai, People's Republic of China

*Corresponding authors: N.Y. (nuo@hust.edu.cn) and B.L. (phylibw@nus.edu.sg)

**Abstract**

The thermoelectric properties of n-type nanoscale three dimensional (3D) Si phononic crystals (PnCs) with spherical pores are studied. Density functional theory and Boltzmann transport equation under the relaxation time approximation are applied to study the electronic transport coefficients, electrical conductivity, Seebeck coefficient and electronic thermal conductivity. We found that the electronic transport coefficients in 3D Si PnC at room temperature (300 K) change very little compared with that of Si, for example, electrical conductivity and electronic thermal conductivity is decreased by 0.26–0.41 and 0.39–0.55 depending on carrier concentration, respectively, and the Seebeck coefficient is similar to that of bulk Si. However, the lattice thermal conductivity of 3D Si PnCs with spherical pores is decreased by a factor of 500 calculated by molecular dynamics methods, leading to the *ZT =0.76* at carrier concentration around $2\times10^{19}$ cm$^{-1}$, which is about 30 times of that of porous Si. This work indicates that 3D Si PnC is a promising candidate for high efficiency thermoelectric materials.

1. Introduction

Thermoelectric materials are essential for electricity generation from waste heat and for cooling.[1] The performance of thermoelectric materials depends on the figure of merit $ZT$,[2] $ZT=S^2\sigma T/\kappa$, where S, T, σ, and κ are the Seebeck coefficient, absolute temperature, electrical conductivity and total thermal conductivity, respectively. Total thermal conductivity κ has contributions from both electrons ($\kappa_e$) and phonons ($\kappa_{ph}$), $\kappa = \kappa_e + \kappa_{ph}$. $ZT$ can be increased by increasing S or σ, or decreasing κ. However, it is difficult to improve $ZT$ in conventional bulk materials. An effective way to increase the value of $ZT$ is to greatly reduce the lattice thermal conductivity with little degradation of electronic properties by nanostructuring.[2]

Si has the advantage of being low-cost, environmentally friendly, and widely used in the semiconductor industry. However, the bulk Si is a poor thermoelectric material because of its high lattice thermal conductivity (~156 W/m-K at 300 K) which results in low value of $ZT$ (~ 0.01 at 300 K).[3] One way to improve the efficiency of Si as TE material is to engineer nanostructured Si by reducing the lattice thermal conductivity.[4] The rough surface Si nanowires has the value of $ZT$ approaching 0.6 at room temperature due to the 100-fold reduction of lattice thermal conductivity.[5] Our previous work found that the lattice thermal conductivity is significantly decreased in nanoscale three-dimensional (3D) Si phononic crystals (PnCs) with spherical pores.[6] For example, $\kappa_{ph}$ of 3D Si PnCs with porosity larger than 90% is decreased by 4 orders of magnitude compared with that of bulk Si at room temperature. Thus, 3D Si PnCs could be expected to have large value of $ZT$.

However, the electronic properties of porous Si with pores randomly arranged both in position and size is substantially deteriorated, which would result in a low value of $ZT$.[7]

Previous work about nanoporous Si with periodic arrangement of cylinder pores have shown that the electronic properties are little degraded and the lattice thermal conductivity is decreased by a factor of 200 compared with that of bulk Si.[8] This results in an enhanced value of *ZT* which is increased by 2 orders of magnitude compared with that of bulk Si with appropriate carrier concentration.[9] Experimental work on holey Si which is constructed by nanopores distributed on Si thin film shows that the value of *ZT* could reach 0.4 at room temperature.[10] However, the electronic properties of 3D Si PnCs with spherical pores has not yet been investigated.

In this work, we study thermoelectric properties of n-type nanoscale 3D Si PnC. The electronic transport coefficients of 3D Si PnCs with spherical pores at 300 K is calculated by a combination of density functional theory (DFT) implemented in Quantum ESPRESSO[11] and the Boltzmann transport equation (BTE) under the relaxation time approximation implemented in BoltzTraP[12]. The lattice thermal conductivity of 3D Si PnC is calculated by equilibrium molecular dynamics (EMD) method. The electronic band structure of 3D Si PnC is calculated to understand the almost unaffected electronic transport coefficients. Furthermore, the porosity effect on electronic transport coefficients is studied.

2. Structure and method

The nanoscale 3D Si PnC is constructed by periodic arrangement of nanoscale supercells, where the supercell is constructed from a cubic cell with a spherical pore (shown in Figure 1). The centers of cubic and spherical pore are overlapped. The period length of nanoscale 3D PnC is the distance between centers of two nearest supercell. The porosity (P) is defined as the ratio of the number of removed atoms in the pore to the total number of atoms in a cubic

Si cell. The relaxed atomic structure is calculated using DFT. For the nanoscale 3D Si PnC in Figure 1, the period length is 3 units (1.63 nm) and the diameter of the pore is 1.0 nm, where 1 unit is 0.543 nm.

The electronic transport coefficients, electrical conductivity σ, Seebeck coefficient S and electronic thermal conductivity $\kappa_e$, can be calculated through Boltzmann transport equation under the relaxation time approximation:[13]

$$\sigma = L^{(0)} \tag{1}$$

$$s = -(1/eT)\sigma^{-1}L^{(1)} \tag{2}$$

$$\kappa_e = 1/(e^2 T)(L^{(2)} - L^{(1)}\sigma L^{(1)}) \tag{3}$$

$$L^{(\alpha)} = e^2 \tau \sum_n \int \frac{d\vec{k}}{4\pi^3} \left(-\frac{\partial f(\epsilon_{n\vec{k}})}{\partial \epsilon_{n\vec{k}}}\right) \vec{v}_{n\vec{k}} \vec{v}_{n\vec{k}} (\epsilon_{n\vec{k}} - \mu)^\alpha \tag{4}$$

where, α is 0, 1 or 2, is the energy eigenvalue of the *n*th band at $\vec{k}$ point in the first Brillouin zone, $f(\epsilon_{n\vec{k}})$ is the Fermi-Dirac distribution function at room temperature *T*, $\mu$ is the chemical potential, $\tau$ is the relaxation time and the group velocity, $\vec{v}_{n\vec{k}} = (1/\hbar)\nabla_{\vec{k}}\epsilon_{n\vec{k}}$. The relaxation time $\tau$ is taken to be energy-independent which is obtained by fitting experimental value of bulk Si.[14] The Seebeck coefficient in Eq. 2 is independent of relaxation time. BoltzTraP is applied here, it is a program for calculating the semi-classic electronic transport coefficients.[12]

The electronic structure of the system is needed for calculating the electronic transport coefficients. All the electronic structure is performed by the density functional theory approach[15,16] as implemented in the Quantum ESPRESSO.[11] The kinetic energy cutoff for wave functions is set as 28 Ry, the kinetic energy cutoff for charge density and potential is set as 280 Ry, ultrasoft Vanderbilt type pseudopotential[17,18] is used, and Perdew-Wang 91 gradient-corrected functional is applied for exchange correlation energy.[19] To find the optimal doping range, the electronic transport coefficients are calculated as a function of carrier concentration with rigid band approximation. To obtain a converge results of the electronic transport coefficients, a dense $\vec{k}$ point mesh with 146 points is used in the irreducible Brillouin zone.

The lattice/phonon thermal conductivity $\kappa_{ph}$ is calculated by equilibrium molecular dynamics simulation through Green-Kubo formula[20], details can be found in the supporting information of Ref. [6]. Stillinger-Weber (SW) potential is applied to describe the interaction between the Si atoms for its accurate fit for experimental results on the thermal expansion coefficients.[21,22] Velocity Verlet algorithm is employed to integrate equations of motion, and the EMD step time is 1.0 fs. Periodic boundary condition is applied in three spatial directions. Initially, Langevin heat reservoir is used to equilibrate the system at 300 K for 1 ns. Then, microcanonical ensemble (NVE) runs for another 16.7 ns. Meanwhile, heat current is recorded at each step. Then, the thermal conductivity is calculated by the Green-Kubo formula. The thermal conductivity is the mean value of twelve realizations with different initial conditions, and the error bar is the deviation of the twelve realizations.

3. Results and discussions

The calculated room temperature electronic transport coefficients as a function of doping concentration ($n_e$) is shown in Figure 2. Figure 2 (d) shows the value of *ZT* as a function of doping concentration. The period length of 3D Si PnC with spherical pores is 1.63 nm, and the porosity of 3D Si PnC is varied by changing the diameter of pore. To compare the electronic transport coefficients of 3D Si PnCs with that of bulk Si, we also calculated the electronic transport coefficients of bulk Si. The red line shows our calculation results of bulk Si and the open circles are the results of bulk Si cited from Lee, Galli and Grossman's work.[9] The results of electronic transport coefficients of bulk Si are consistent with each other.

Compared with that of bulk Si, the electrical conductivity of 3D Si PnC with porosity 30.5% is decreased from 26% to 41% (Figure 2 (a)) depending on the carrier concentration. Additionally, the electrical conductivity is reduced as the porosity increases; however, the reduction is not significant. As shown in Figure 2 (b), the electronic thermal conductivity of 3D Si PnC with porosity 30.5% is decreased by from 39% to 55% compared with that of bulk Si. The reduction of electrical conductivity and the electronic thermal conductivity is due to the porous structure of Si PnC, because there are more boundary atoms on the surface of pores which cause more scatterings of electrons. The Seebeck coefficient of 3D Si PnC is close to that of bulk Si when carrier concentration is larger than $2\times10^{19}$cm$^{-3}$ (Figure 2 (c)). Interestingly, the Seebeck coefficient of 3D Si PnCs changes little as the porosity changes. These results indicate that the electronic properties can be preserved in nanoscale 3D Si PnC.

The electronic band structure and density of states (DOS) of bulk Si and 3D Si PnC with spherical pores are shown in Figure 3 (a) and Figure 3 (b), respectively. For comparison, the same tetragonal symmetry is used in the calculation of electronic band structure of bulk Si and 3D Si PnC. The period length is 1.63 nm (3 units), and the diameter of the pores is 1.0 nm for 3D Si PnC. The electronic band gap of bulk Si is 0.57 eV, whereas the electronic band gap of 3D Si PnC is 1.01 eV. The electronic band gap is broadened compared with that of bulk Si. Furthermore, the band is flattened due to the periodic pore along the Γ to X direction, which could result in a reduction of electrical conductivity and electronic thermal conductivity. However, the band structure along other directions such as Γ to M, M to X and Γ to R of 3D Si PnC remains as dispersive as that in the bulk Si, thus the reduction of electrical conductivity and electronic thermal conductivity is not significant as the lattice thermal conductivity which will be discussed later. The density of states of Si PnC is larger than that of Si near Fermi level which is consistent with the flattening of the dispersions.

The calculated lattice thermal conductivity, $\kappa_{ph}$, (dash lines in Figure 2 (b)) of Si PnC with porosity 13.9%, 30.5%, 38% are 1.30±0.01, 0.50±0.01, 0.36±0.01 W/m-K at room temperature, respectively. The lattice thermal conductivity of bulk Si is 170±16 W/m-K by EMD method. Compared with bulk Si, the lattice thermal conductivity of 3D PnC is decreased by a factor up to 500. As shown in our previous work,[6] the extreme low thermal conductivity of 3D Si PnC is caused by the great localization of phonon modes. Because there are more phonon localizations in 3D Si PnC with larger porosity, the thermal conductivity decreases as the porosity increases.[6] Additionally, the phonon group velocities are decreased in 3D Si PnC, which could also result in a reduction of thermal conductivtiy.[6] Generally, the lattice thermal conductivity dominate in the contribution of heat transfer of semiconductor and dielectrics. Differently, the lattice thermal conductivity of 3D Si PnC is decreased to the same order of magnitude as that of electronic thermal conductivity (solid line in Figure 2 (b)), which may make a great enhancement of the value of *ZT*.

As shown in Figure 2 (d), the value of *ZT* in 3D Si PnC is greatly increased by around 100 times compared with bulk Si. The value of *ZT* is enhanced to 0.76 when the porosity is 30.5% at the carrier concentration around $2\times10^{19}$ cm$^{-3}$. When the porosity is increased to 38%, although the lattice thermal conductivity is decreased, the value of *ZT* is reduced due to a larger decrease of electronic transport coefficients in 3D Si PnC. In Table 1, we list the value of *ZT* of different Si structures and high *ZT* thermoelectric materials. The values of *ZT* of Si nanostructures are greatly increased compared with that of bulk Si. In addition, *ZT* of 3D Si PnC is the largest in all Si nanostructures, although it is lower than that of other complex materials, like $Bi_2Te_3/Sb_2Te_3$ and $In_{0.53}Ga_{0.47}As$ whose value are larger than 1.0. Si nanostructures have advantage of easy fabrication, low cost and widely used in semiconductor industry. The calculation results suggest that nanoscale 3D Si PnC is a good candidate for the future thermoelectric materials.

4. Conclusion

In this work, we studied the thermoelectric properties of 3D Si PnC with spherical pores. We found that the electrical conductivity and electronic thermal conductivity is decreased little compared with that of bulk Si, and the Seebeck coefficient is close to that of bulk Si. The electronic band structure of 3D Si PnC is as dispersive as that of bulk Si, which could result in the little deterioration of electronic transport coefficients. The lattice thermal conductivity is decreased by a factor up to 500 compared with that of bulk Si using EMD method. By calculating the value of *ZT*, we found that 3D Si PnC with optimized carrier concentration could have *ZT* reaching 0.76. The calculation results of this work suggest that nanoscale 3D Si PnC is a promising candidate for thermoelectric materials.

**Acknowledgments**

B.L. was supported in part by MOE Grant R-144-000-305-112 of Singapore, and the National Natural Science Foundation of China (11334007). N.Y. was supported in part by the grants from Self-Innovation Foundation (2014TS115) of HUST, Talent Introduction Foundation (0124120053) of HUST, and the National Natural Science Foundation of China Grant (11204216). We are grateful to Prof. Chun Zhang, and Dr. Liyan Zhu for useful discussions.

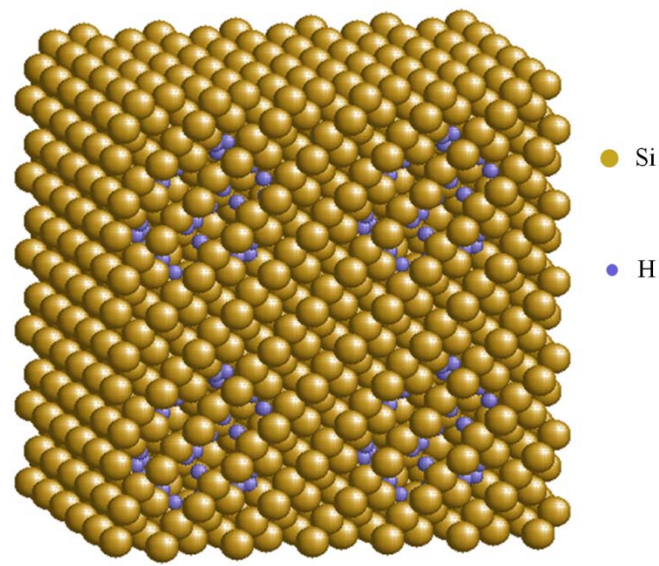

Figure 1. Structure of Si 3D PnCs with spherical pores. The Si atoms in the internal surface are passivated with hydrogen atom H. The yellow atoms are Si, and the purple atoms are H. The period length is 1.63 nm, the diameter of the spherical hole in this 3D Si PnC is 1.0 nm, and the porosity is calculated as 13.9%. Here 1 unit is 0.543 nm.

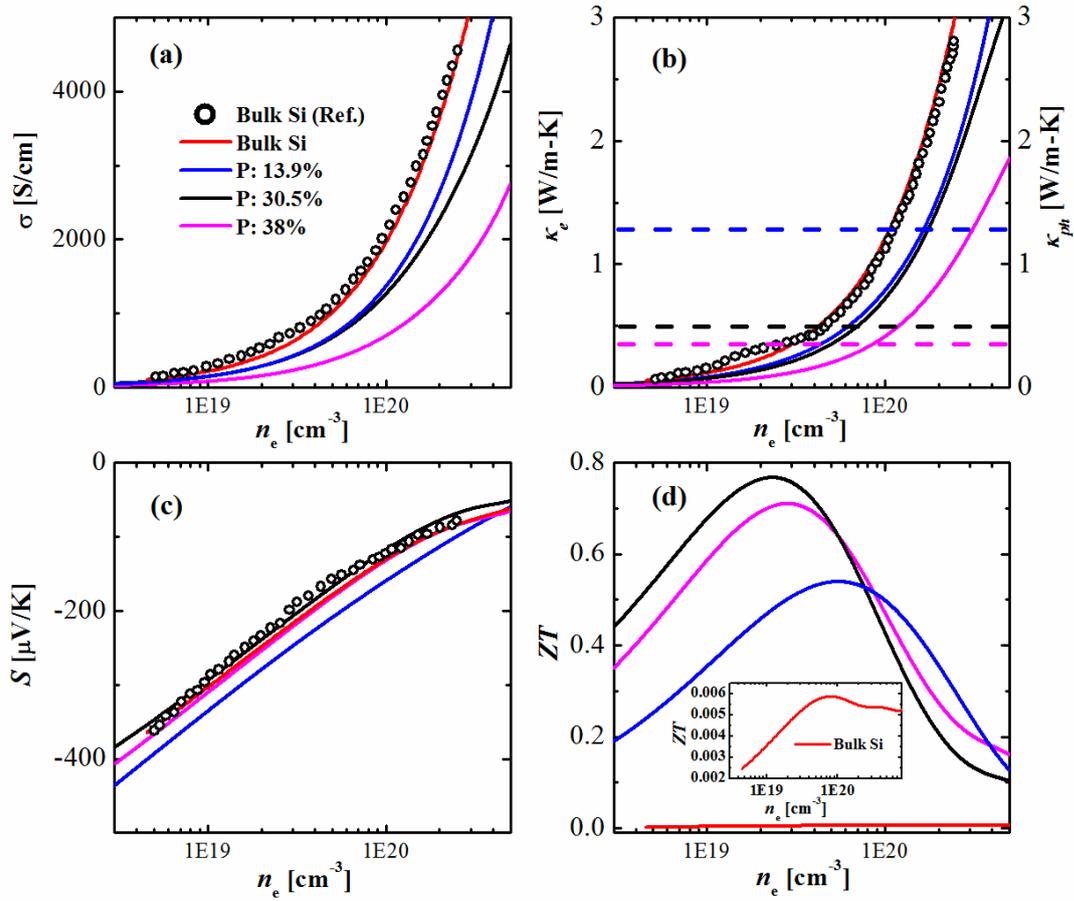

Figure 2. The electronic transport coefficients of bulk Si and Si PnCs as a function of doping concentration ($n_e$) at room temperature. (a) Electrical conductivity; (b) Electronic thermal conductivity $\kappa_e$ (solid line) and calculated lattice thermal conductivity $\kappa_{ph}$ by EMD method (dash line); (c) Seebeck coefficient; (d) Figure of merit of $ZT$. The red lines correspond to the electronic transport coefficients of bulk Si, which is close to the results of bulk Si from Ref. [9] (the open circles). The other three color lines correspond to 3D Si PnC with the porosity (P) of 13.9%, 30.5% and 38%, correspondingly. The three dash lines in (b) correspond to the lattice thermal conductivity $\kappa_{ph.}$ of 3D Si PnC with PnC with the porosity (P) of 13.9%, 30.5% and 38%, respectively. The period length of the 3D Si PnC with spherical pores is 1.63 nm, and the porosity is changed by varying the diameter of pore.

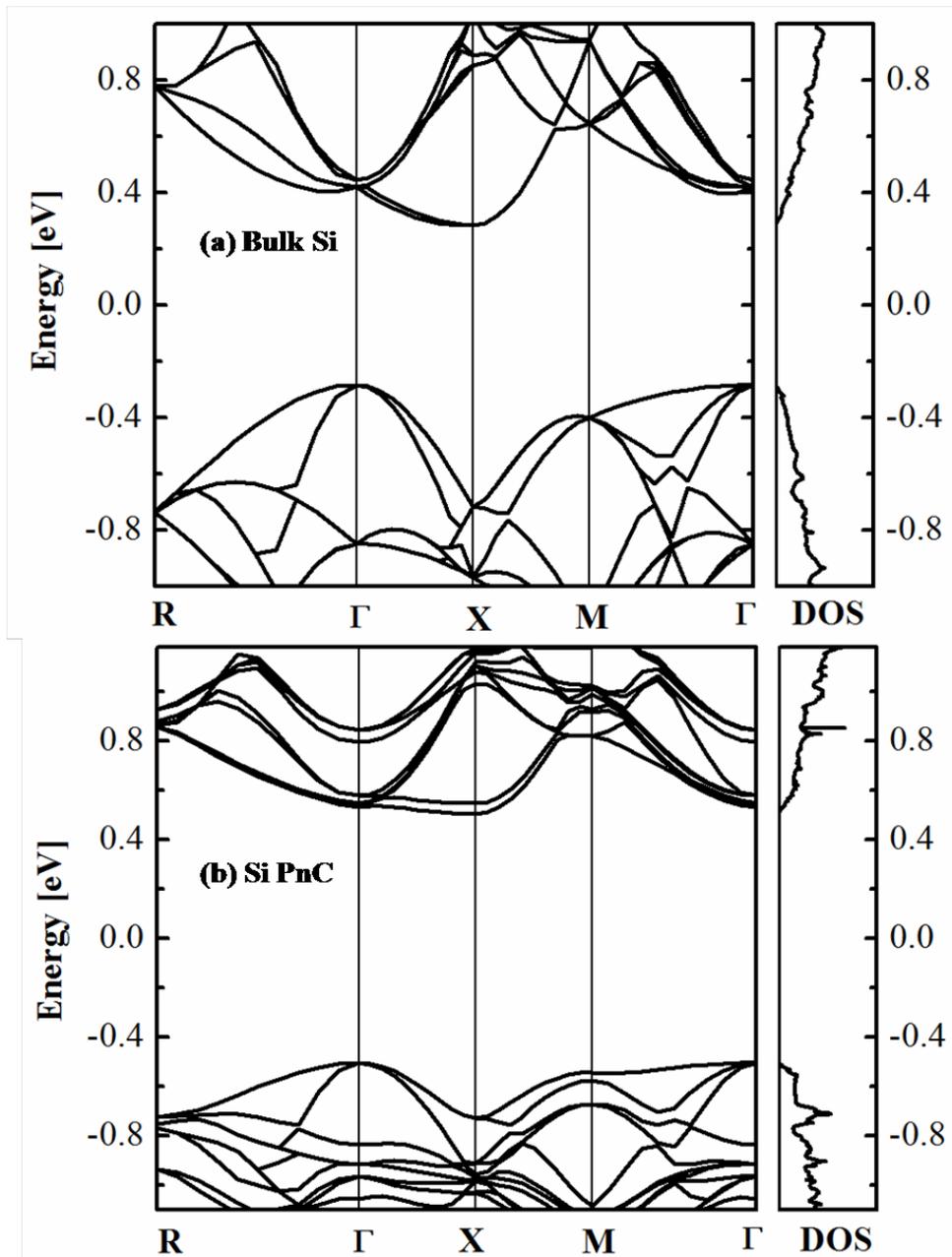

Figure 3 (a) The electronic band structure and density of states (DOS) of bulk Si. (b) The electronic band structure and DOS of 3D Si PnC with spherical pores. In the calculation of the band structure of both bulk Si and 3D Si PnC, the same tetragonal symmetry is used for comparison. Period length of 3D Si PnC is 1.63 nm (3 units) and diameter of the pore is 1 nm.

Table 1. The value of figure of merit (*ZT*) of different materials. NEGF represents the nonequilibrium Green function methodology.

| Material | Method | Temperature | *ZT* |
|---|---|---|---|
| Bulk Si | DFT+BTE+MD | 300 K | 0.006 |
| **3D Si PnC with P of 30.5%** | **DFT+BTE+MD** | **300 K** | **0.76** |
| Si Nanoporous[9] | DFT+BTE+MD | 300 K | 0.4 |
| Ge Nanoporous[23] | DFT+BTE+MD | 300 K | 0.83 |
| Porous Si[24] | Experiment | 310 K | 0.025 |
| Holey Si[10] | Experiment | 300 K | 0.4 |
| Si Nanowires[4] | Experiment | 200 K | 1.0 |
| Si Nanowires[4] | Experiment | 300 K | 0.4 |
| Rough Si Nanowires[5] | Experiment | 300 K | 0.6 |
| Nanostructured SiGe bulk alloy[25] | Experiment | 300 K | 0.52 |
| Nanostructured SiGe bulk alloy[25] | Experiment | 900 K | 1.3 |
| Graphene nanoribbons with heavy adatoms and nanopores[26] | NEGF+first principle method | 40 K | 3 |
| $Bi_2Te_3/Sb_2Te_3$ thin film superlattice[27] | Experiment | 300 K | 2.4 |
| Embedding nanoparticles in $In_{0.53}Ga_{0.47}As$[28] | Experiment | 300 K | 2 |
| SnSe[29] | Experiment | 300 K | 0.2 |
| SnSe[29] | Experiment | 923 K | 2.6 |
| PbTe-SrTe doped with Na[30] | Experiment | 300 K | 0.1 |
| PbTe-SrTe doped with Na[30] | Experiment | 915 K | 2.2 |
| Pb-Te-CdTe alloy[31] | Experiment | 775 K | 1.7 |
| Nanostructured In-doped SnTe[32] | Experiment | 300 K | 0.1 |
| Nanostructured In-doped SnTe[32] | Experiment | 873 K | 1.1 |